\documentclass[prl,twocolumn,showpacs]{revtex4}
\usepackage{graphicx}

\newcommand{\pv}[2]{{#1}^{(\!#2\!\!\>)}}
\newcommand{\refeq}[1]{(\ref{#1})}
\newcommand{\reffig}[1]{Figure \ref{#1}}

\newcommand{\od}[2]{\frac{d{#1}}{d{#2}}}

\begin{document}

\title{Self-Organized Bottleneck in Energy Relaxation}
\author{Hidetoshi Morita}
\email{morita@complex.c.u-tokyo.ac.jp}
\author{Kunihiko Kaneko}
\affiliation{
Department of Pure and Applied Sciences, University of Tokyo,
Komaba, Meguro-ku, Tokyo 153-8902, Japan
}
\date{\today}

\begin{abstract}
We study an energy relaxation process
after many degrees of freedom are excited
in a Hamiltonian system with a large number of degrees of freedom.
Bottlenecks of relaxation,
where relaxations of the excited elements are drastically slowed down,
are discovered.
By defining an internal state for the excited degrees of freedom,
it is shown that the drastic slowing down occurs
when the internal state is in a critical state.
The relaxation dynamics brings the internal state into the critical state,
and the critical bottleneck of relaxation is self-organized.
Relevance of our result to relaxation phenomena
in condensed matters or large molecules
is briefly discussed.
\end{abstract}

\pacs{05.70.Ln, 05.45.-a, 87.10.+e}
\maketitle

Relaxation process to equilibrium
has gathered much attention of physicists over several decades.
Although the relaxation near equilibrium
has been well formulated theoretically,
the process far from equilibrium is not yet fully understood,
while relaxation phenomena after strong excitation
can show novel and interesting behaviors.
For example, single-molecule Myosin, a motor protein,
shows anomalously long energy storage
when biochemically excited by ATP~\cite{Ishijima}.
Such slow relaxations are also found
in the dynamics of single enzymes~\cite{Lu,Edman-Rigler}.

When a system is excited weakly,
it is described by a superposition of elementary excitations
with weak interactions.
When the excitation is strong, on the other hand,
it is inevitable to take into account strong interactions
of the excited modes, which may behave cooperatively.
Even though a general theory for such case may not be available,
it is important to find a class of novel relaxation phenomena, 
and propose novel physical concepts associated with it.

In this Letter,
we report a relaxation process
when quite a few degrees of freedom are highly excited,
in a simple Hamiltonian system with large degrees of freedom.
The excited elements which interact strongly form 
a partial thermodynamic system, to which an internal state
(with effective temperature etc.) is assigned.
The relaxation depends on this internal state,
and, in turn, the internal state dynamically changes with the relaxation.
We will show that,
due to the interplay between the relaxation and the internal state dynamics,
the partial system self-organizes a critical state,
which forms a bottleneck in the relaxation course.
We discuss its mechanism,
in particular by emphasizing distinction of it
from critical slowing down and self-organized criticality~\cite{SOC}.

As a specific model
we adopt a Hamiltonian system
~\cite{Konishi-Kaneko,Ruffo};
\begin{equation}
H(\theta,p)=\sum_{i=1}^{N} \frac{{p_i}^2}{2}
+ \frac{1}{2N} \sum_{i=1}^{N}\sum_{j=1}^{N} V(\theta_i-\theta_j),
\label{eq:Hamiltonian}
\end{equation}
with,
\begin{equation}
V(\theta)=\frac{K}{(2\pi)^{2}}\left[ 1-\cos2\pi\theta \right].
\end{equation}
The equations of motion are given by,
\begin{equation}
\od{p_i}{t} = -\frac{K}{2\pi N}\sum_{j=1}^{N} \sin2\pi (\theta_i-\theta_j),
\end{equation}
where $N$ pendula are globally coupled,
interacting each other through phase difference $2\pi(\theta_i-\theta_j)$.
Each pendulum can have two types of motion;
rotation at a higher energy and libration at a lower energy.

This system is known to exhibit continuous phase transition in equilibrium
in the thermodynamic limit $N\to\infty$~\cite{Ruffo}.
Order parameter $M$ for this transition is defined as,
\begin{math}
Me^{i2\pi\phi}=\sum_{j=1}^{N}e^{i2\pi\theta_j}/N.
\end{math}
When total energy $E$ is very small, $M\simeq 1$, where all pendula librate.
As the total energy increase, $M$ decrease,
where some pendula librate and others rotate,
changing their types of motion temporally.
Above the transition energy $E_c=0.75/(2\pi)^2$, $M\simeq 1/\sqrt{N}$,
and all pendula rotate almost freely.

Energy relaxation when a single pendulum is excited
is found to be very slow~\cite{Nakagawa-Kaneko},
with the relaxation time increasing
as a stretched exponential form of the excited energy,
as is discussed in terms of Boltzmann-Jeans conjecture~\cite{BJC,BJC_recent}.

Here, we are interested in relaxation properties
when a part of the system composed of many pendula are excited. 
At $t=0$, $\pv{N}{K}$ pendula are simultaneously kicked,
i.e. excited instantaneously, with the same momentum $\pv{P}{K}$.
Here, we consider the case that the rest system is huge enough,
and is regarded to be in thermal equilibrium.  
Instead of carrying out numerical simulations of the Hamiltonian system
\refeq{eq:Hamiltonian} with such huge degrees of freedom,
we add a heat bath to the rest of the system;
i.e. equations of the motion of the rest $N-\pv{N}{K}$ pendula
are given by the Langevin equation,
\begin{equation}
\od{p_i}{t} = -\frac{K}{2\pi N}\sum_{j=1}^{N} \sin2\pi (\theta_i-\theta_j)
-\gamma p_i +\xi_i(t),
\end{equation}
where $\xi_i(t)$ is white-gaussian random noise,
that is $\langle\xi_i(t)\rangle =0$
and $\langle\xi_i(t_0)\xi_j(t_0+t)\rangle =2\gamma T\delta_{ij}\delta(t)$.

\begin{figure}
\includegraphics[scale=0.4]{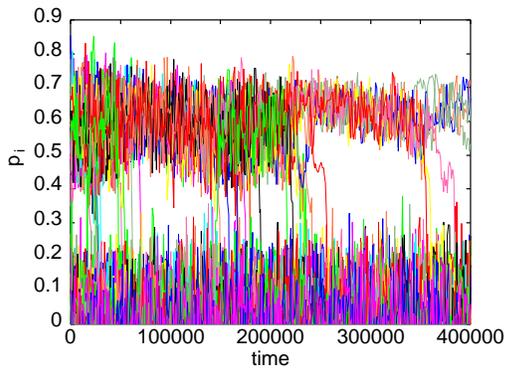}
\caption{
A typical time series of momenta of pendula.
$N=64, T=0.01, \pv{N}{K}=32, \pv{P}{K}=0.6$.
\label{fig:p_tseri}
}
\end{figure}

\begin{figure}
\includegraphics[scale=0.4]{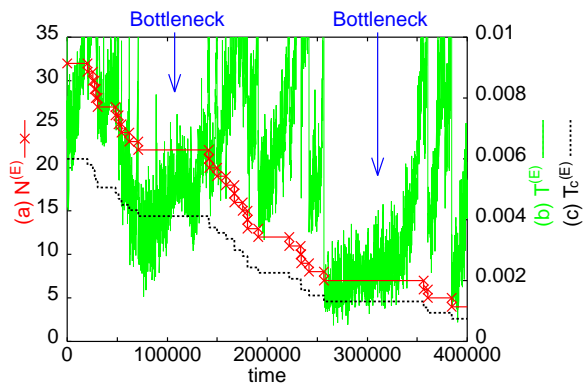}
\caption{
A typical time series of 
(a) the population of excited pendula $\pv{N}{E}$,
(b) the effective temperature of the excited part $\pv{T}{E}$
(\refeq{eq:Te_def}),
and
(c) the effective critical temperature ${\pv{T_c}{E}}$
~\cite{scaling}.
The data are the same as \reffig{fig:p_tseri}.
\label{fig:Te_tseri}
}
\end{figure}

A typical time series of the energy relaxation after the kick
is drawn in \reffig{fig:p_tseri}.
The excited pendula that rotate relax one by one
to librate with the rest of the pendula,
so that the \textit{population of excited pendula} $\pv{N}{E}$
decrease from $\pv{N}{K}$ to zero,
whose time series is drawn in \reffig{fig:Te_tseri}-(a).
It shows several plateaus, where the relaxation is drastically slowed down.
We call these plateaus as \textit{bottlenecks} of the relaxation.

With regards to this intermittent appearance of the relaxation bottlenecks,
we ask whether the bottleneck appears randomly or follows some rule. 
If the latter is the case,
the state of the excited partial system should be important,
since other huge degrees are kept at a constant temperature
during the relaxation.

Here,
the interaction between the rotating and librating pendula
is small on the average,
because the temporal average of the term
$\sin 2\pi(\theta_i(t)-\theta_j(t))$ is negligible
if the momenta of the two pendula are much different.
Indeed, this is the origin of long relaxation time scale
in the Boltzmann-Jeans conjecture~\cite{Nakagawa-Kaneko}.
The interaction between two excited pendula, on the other hand, can be strong,
if the momenta are not much different.
In fact, by taking a Galilean frame
moving with the average velocity of the excited pendula,
they often librate each other, keeping attracting interactions.

Hence our system is represented by two partial systems
which are weakly coupled.
One system is composed of excited pendula
whose momenta $\simeq \pv{P}{E}$,
while the rest system has momenta $\simeq 0$.
Since the interaction between the two partial systems are small,
it is relevant to consider a thermodynamic state of each partial system
at the first-order approximation.
The state of the relaxed part is basically determined
by the heat bath with the temperature $T$.
The state of the excited part, on the other hand,
changes as the relaxation process progresses.
Thus it is important to study a thermodynamic state of the excited part
as an \textit{internal state} of the whole system.

As a quantity representing the internal state,
we introduce \textit{effective temperature of the excited part} $\pv{T}{E}$ as
\begin{equation}
\pv{T}{E}=\frac{1}{\pv{N}{E}}\pv{\sum_i}{E}(p_i - \pv{P}{E})^2,
\label{eq:Te_def}
\end{equation}
by taking an inertial frame
with the center-of-mass momentum of the excited part
$\pv{P}{E}=\pv{\sum_i}{E}p_i/{\pv{N}{E}}$,
where $\pv{\sum}{E}$ denotes the summation
over excited pendula~\cite{partial_quantity}.

\begin{figure}
\includegraphics[scale=0.4]{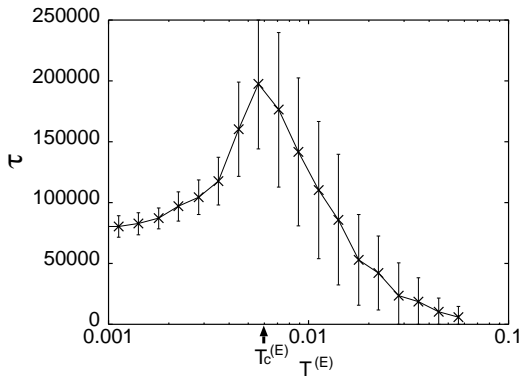}
\caption{
The first relaxation time $\tau$
versus the effective temperature of excited part $\pv{T}{E}$.
$\pv{T}{E}$ is computed from the time average of \refeq{eq:Te_def}
over a period of 1000 just after the pendula are excited~\cite{period1000},
where the time scale of libration is the order of 0.1 to 1.
Instead of simple kick, we here attach two systems at $t=0$,
whose center-of-mass momenta are $\pv{P}{K}$ and 0.
By varying the temperature at $t<0$, $\pv{T}{E}$ is controlled.
$N=32$, $T=0.01$, $\pv{N}{K}=16$, $\pv{P}{K}=0.8$.
2000 samples are computed in all.
$\tau$ is averaged for the samples within the bin size of 0.1 for log scale.
$\pv{T_c}{E}$ is the effective critical temperature
defined as $\pv{T_c}{E}=T_c\pv{N}{E}/N$~\cite{scaling}.
\label{fig:tau_vs_Te}
}
\end{figure}

To see how the relaxation depends on the internal state,
we first define the \textit{first relaxation time} $\tau$ as the time
when the first one of $\pv{N}{K}$ excited pendula loses energy to librate.
In other words, $\tau$ is the first escape time of a pendulum
from the excited to the relaxed part.
The relation between $\tau$ and $\pv{T}{E}$
is plotted in \reffig{fig:tau_vs_Te},
where $\pv{T}{E}$ is time-averaged
for a while just after the pendula are excited.
A peak is discernible.
The temperature for this peak corresponds to the
\textit{critical} point for the phase transition 
of the excited part $\pv{T_c}{E}=T_c \pv{N}{E} /N$~\cite{scaling},
when the excited part is regarded as in equilibrium.

We have also computed the time series of
$\pv{T}{E}$ and $\pv{T_c}{E}$,
and plotted in \reffig{fig:Te_tseri}-(b) and (c), respectively.
The plateau of $\pv{N}{E}$ actually corresponds to the time
when the effective temperature of the excited pendula is
around the critical temperature.
From these results it is concluded that the relaxation bottleneck occurs
when the internal state is around a critical point.

One might think that this phenomenon is
nothing but a critical slowing down of the excited system.
This is not the case.
First note that the whole system is not necessarily in a critical state.
In fact we verified
that the coincidence of this peak position to the effective critical point
is qualitatively independent of the heat bath temperature $T$ (not shown).
Next note that such long relaxation bottleneck is not found
when disregarding the rest system.
In fact we estimated the first escape time of a pendulum
in only a single system,
by calculating the time until one of the pendula
first takes a momentum smaller than a threshold.
This escape time increases with the decrease of the temperature,
without showing any peak (not shown).
Hence the present slow relaxation is \textit{inter}-system relaxation,
not \textit{intra}-system relaxation
to which the so-called critical slowing down belongs.

\begin{figure}
\includegraphics[scale=0.4]{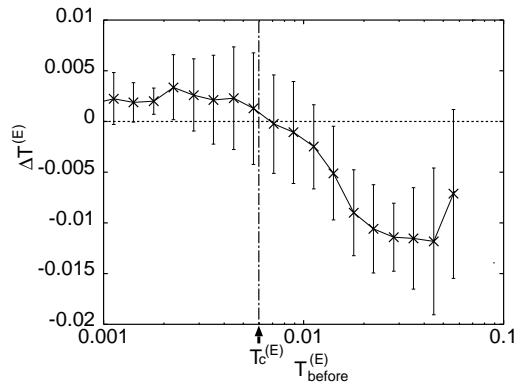}
\caption{
$\Delta\pv{T}{E}$ versus $\pv{T}{E}_{before}$,
computed from the simulations of \reffig{fig:tau_vs_Te}.
$\pv{T}{E}_{before}$ and $\pv{T}{E}_{after}$ are computed
from the time average over a period of 1000 just before and after $\tau$~\cite{period1000},
respectively.
$\pv{\Delta T}{E}$ is averaged for the samples
within the bin size of 0.1 for log scale.
$\pv{T_c}{E}$ is the effective critical temperature.
\label{fig:dTe_vs_Te}
}
\end{figure}

Then how is the excited system sticked to the critical point
through  the interaction with the rest system? 
To answer this question, we consider how the internal state changes 
before and after one of the excited pendula relaxes.
To see this change,
we introduce \textit{effective temperature change} $\Delta\pv{T}{E}$
defined as,
\begin{equation}
\Delta\pv{T}{E}
\equiv \frac{\pv{N}{E}}{\pv{N}{E}-1}\pv{T}{E}_{after}
- \pv{T}{E}_{before},
\label{eq:Delta_Te_DEF}
\end{equation}
where $\pv{T}{E}_{before}$ and $\pv{T}{E}_{after}$ is the effective temperature
before and after the first relaxation of an excited pendulum, respectively,
while the factor $\pv{N}{E}/(\pv{N}{E}-1)$
comes from the contribution of scaling~\cite{scaling}.
The relation between $\pv{\Delta T}{E}$ and $\pv{T}{E}_{before}$
is plotted in \reffig{fig:dTe_vs_Te},
where $\pv{T}{E}_{before}$ and $\pv{T}{E}_{after}$ are
time averages over a given period just before and after $\tau$, respectively.
The average of $\Delta\pv{T}{E}$
is positive when $\pv{T}{E}_{before}<\pv{T}{E}_c$,
while negative when $\pv{T}{E}_{before}<\pv{T}{E}_c$.
This result indicates
that the internal state tends to be attracted into the critical state,
through the interaction between the excited system and the rest.

To analyze the above temperature dependence of $\Delta\pv{T}{E}$,
we note two factors for it;
the increase of the temperature
by work given to the excited part from the relaxed part,
and the decrease of the temperature
by an escape of a pendulum from the excited part.

\begin{figure}
\includegraphics[scale=0.4]{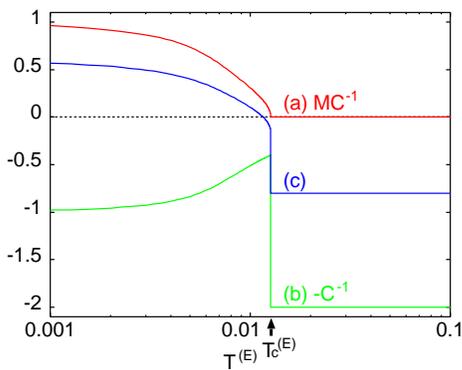}
\caption{
Temperature dependences of (a)$MC^{-1}$ and (b)$-C^{-1}$
in the thermodynamic limit $N\to\infty$,
which are obtained analytically
by considering canonical distribution~\cite{Ruffo}.
In the present mean-field model,
the critical exponent of $M$ at $T=T_c-0$ is $1/2$,
and so the specific heat $C$ does not diverge but takes a largest finite value.
(c) is an example of combination of the two factor (a) and (b)
with the ratio of 1:0.4.
The temperature of horizontal axis is for a single system,
and has to be scaled with $\pv{N}{E}/N$
to compare with \reffig{fig:dTe_vs_Te}.
\label{fig:analysis}
}
\end{figure}

First,
a work is given from the relaxed part to the excited part,
which causes temperature increase
as the second law of thermodynamics indicates.
Temperature change by the work is given by
$\Delta\pv{T}{E}=\int dt\;\pv{\dot{T}}{E}(t)$
with
\begin{equation}
\pv{\dot{T}}{E}
=\frac{2}{\pv{N}{E}}\pv{\sum_j}{E}p_j\dot{p_j}
-2\pv{P}{E}\pv{\dot{P}}{E}.
\end{equation}
Using the order parameter of the partial systems
\begin{math}
\pv{M}{*}e^{i\pv{\phi}{*}}=\pv{\sum_j}{*}e^{i\theta_j}/\pv{N}{*},
\end{math}
where $*$ is $E$ or $R$~\cite{partial_quantity},
we obtain the following expression,
which is closed only by macroscopic variables:
\begin{eqnarray}
\lefteqn{\frac{d}{dt}\pv{E}{E} = K\frac{\pv{N}{R}}{N}\left\{
\frac{d}{dt}\left(\pv{M}{E}e^{i\pv{\phi}{E}}\right)\right.}\hspace{25mm} \nonumber\\
&&\left.+i\pv{M}{E}e^{i\pv{\phi}{E}}\pv{P}{E}
\right\}\pv{M}{R}e^{-i\pv{\phi}{R}}.
\label{eq:dE}
\end{eqnarray}
Here $\pv{E}{E}$ is the effective total energy of the excited part,
given by
\begin{math}
\pv{E}{E} = \frac{\pv{T}{E}}{2} +
\frac{\pv{N}{E}}{N}\frac{K}{2}\left[1-{\pv{M}{E}}^2\right].
\end{math}
The first term of the r.h.s. of \refeq{eq:dE} is a fast varying term,
and can be neglected.
With this approximation,
the total energy change $\pv{E}{E}$ of the excited part
is proportional to $\pv{M}{E}$.
Since $\pv{\Delta T}{E}\simeq\pv{\Delta E}{E}/\pv{C}{E}$,
where $\pv{C}{E}$ is effective specific heat of the excited part,
we obtain,
\begin{equation}
\pv{\Delta T}{E} \propto \pv{M}{E}/\pv{C}{E}.
\end{equation}
Its temperature dependence in the thermodynamic limit of the excited part
is drawn in \reffig{fig:analysis}-(a),
which is positive at $\pv{T}{E}<\pv{T_c}{E}$
and vanishes for $\pv{T}{E}>\pv{T_c}{E}$.

Second,
the temperature of the excited part decreases
when a pendulum escapes from the excited part,
which has the largest negative momentum
at the center-of-mass inertial frame of the excited part.
This factor is estimated by calculating the decrease of temperature
with the decrease of energy,
by neglecting the interaction with the relaxed part.
The escape of a pendulum,
with momentum $p_{esc}$ measured from
the center-of-mass inertial frame,
takes away the energy $\Delta E_{esc}=p_{esc}^2/2$ from the excited part.
Since temperature is a function of the total energy
as $\pv{T}{E}=\pv{T}{E}(\pv{E}{E})$,
the effective temperature change \refeq{eq:Delta_Te_DEF} can be described as,
$\Delta\pv{T}{E}=
\pv{N}{E}/(\pv{N}{E}-1)\pv{T}{E}(\pv{E}{E})
-\pv{T}{E}(\pv{E}{E}-\Delta E_{esc})$.
In the limit $\pv{N}{E}\to\infty$,
this is estimated by the specific heat $\pv{C}{E}$ as,
\begin{equation}
\Delta\pv{T}{E}\simeq-{\pv{C}{E}}^{-1}\Delta E_{esc} \propto -{\pv{C}{E}}^{-1},
\end{equation}
whose temperature dependence is drawn in \reffig{fig:analysis}-(b).

By combining above the two factors,
\reffig{fig:dTe_vs_Te} can be explained qualitatively,
as shown in \reffig{fig:analysis}-(c).
$\Delta\pv{T}{E}$ is positive at a temperature below $\pv{T_c}{E}$.
As the temperature approaches $\pv{T_c}{E}$ from below,
$\Delta\pv{T}{E}$ decreases,
Above $\pv{T_c}{E}$,
there is only the negative contribution.
Hence
$\Delta\pv{T}{E}$ changes its sign from positive to negative
as the temperature is increased from below to above $\pv{T_c}{E}$.

In summary, 
by using simple Hamiltonian system,
we have discovered a bottleneck of energy relaxation,
given by a critical state of an internal state of the excited part.
This critical state is self-organized through the course of relaxation.
The long-term relaxation from highly excited state to equilibrium
consists of alternation
between approach to critical state and collapse of the state.
In contrast to the so-called self-organized criticality~\cite{SOC},
the critical state is \textit{spontaneously} formed without continuous driving,
once a \textit{part} of a system is highly excited.
Although we have demonstrated this bottleneck critical state
by a simple Hamiltonian system,
the mechanism is expected to be general,
as long as the system can exhibit continuous phase transition,
and one part of the system is highly excited.

The self-organized bottleneck at a critical state
may be observed in condensed matter,
and in molecules with sufficiently large degrees of freedom,
when a partial system is highly excited.
One candidate for the experimental verification
is partial excitation of semi-conductor material by laser radiation,
where so-called exciton Mott transition is found~\cite{Gonokami}.
Also proteins show drastically slow relaxation
when they are functioning as motors~\cite{Ishijima}
or enzymes~\cite{Lu,Edman-Rigler}.
The present bottleneck may shed new light
on the understanding of such phenomena.

The authors are grateful to S. Sasa and N. Nakagawa for discussion,
and A. Shimizu for informing Ref~\cite{Gonokami}.
This work was supported by a Grant-in-Aid for Scientific Research
from the Ministry of Education, Science, and Culture of Japan.


\begin{thebibliography}{99}
\bibitem{Ishijima}
A. Ishijima, et.al., Cell \textbf{92}, 161 (1998)

\bibitem{Lu}
H. P. Lu, L. Xun, and X. S. Xie, Science \textbf{282}, 1877 (1998)

\bibitem{Edman-Rigler}
L. Edman and R. Rigler, Proc. Natul. Acad. Sci. USA \textbf{97}, 8266 (2000)

\bibitem{SOC}
P. Bak, C. Tang, and K. Wiesenfeld, Phys. Rev. Lett. \textbf{59}, 381 (1987)

\bibitem{Konishi-Kaneko}
T. Konishi and K. Kaneko, J. Phys. A, \textbf{25}, 6283 (1992)

\bibitem{Ruffo}
M. Antoni and S. Ruffo, Phys. Rev. E \textbf{52}, 2361 (1995);
V. Latora, A. Rapisarda, and S. Ruffo, Phys. Rev. Lett. \textbf{83}, 2104 (1999);
Y. Y. Yamaguchi, Prog. Theor. Phys. \textbf{95}, 717 (1996)

\bibitem{Nakagawa-Kaneko}
N. Nakagawa and K. Kaneko, J. Phys. Soc. Jpn. \textbf{69}, 1255 (2000);
Phys. Rev. E \textbf{64}, 055205(R) (2001)

\bibitem{BJC}
L. Boltzmann, Nature \textbf{51}, 413 (1895);
J. H. Jeans, Philos. Mag. \textbf{6}, 279 (1903); \textit{ibid}. \textbf{10}, 91 (1905);
L. D. Landau and E. Teller, Physik. Z. Sowjetunion \textbf{11}, 18 (1936)

\bibitem{BJC_recent}
G. Benettin, L. Galgani, and A. Giorgilli, Commun. Math. Phys. \textbf{121}, 557 (1989);
O. Baldin and G. Benettin, J. Stat. Phys. \textbf{62}, 201 (1991)

\bibitem{partial_quantity}
$\pv{\cdot}{E}$ and $\pv{\cdot}{R}$ stand for
effective macroscopic quantities of the excited and the relaxed part,
respectively,
excepting that $\pv{\sum}{E}$ and $\pv{\sum}{R}$ stand for 
summation over the elements of each part,
respectively.

\bibitem{period1000}
In case $\tau$ ( or the second relaxation time $\tau_2$ )
is smaller than 1000,
the time average is calculated over $\tau$ ( or $\tau_2$ ),
instead of 1000.

\bibitem{scaling}
Hamiltonian \refeq{eq:Hamiltonian} is scaled with $K$,
by $(\theta,p,t,H) \to (\theta,\sqrt{K}p,t/\sqrt{K},KH)$.
Each element evolves
by the equation of motion with factor $K/N$.
Hence the excited part as a thermodynamic system
is scaled with the effective coupling constant $\pv{K}{E}$
defined as $K/N=\pv{K}{E}/\pv{N}{E}$.
Since the temperature is also scaled as $T\to KT$,
the effective critical temperature $\pv{T_c}{E}=T_c\pv{K}{E}/K=T_c\pv{N}{E}/N$,
which decreases as $\pv{N}{E}$ decreases.

\bibitem{Gonokami}
M. Nagai, R. Shimano, and M. Kuwata-Gonokami, J. Lumin. \textbf{87}-\textbf{89}, 192 (2000)
\end{thebibliography}
\end{document}